# Molecular Motor-Powered Shuttles along Multi-walled Carbon Nanotube Tracks

Aurélien Sikora,[1] Javier Ramón-Azcón,[1] Kyongwan Kim,[1] Kelley Reaves,[2] Hikaru Nakazawa,[3] Mitsuo Umetsu,[3] Izumi Kumagai,[3] Tadafumi Adschiri,[1] Hitoshi Shiku,[1,4] Tomokazu Matsue,[1,4] Wonmuk Hwang[2,5,6] and Winfried Teizer*[,1,2,6]

[1] WPI Advanced Institute for Materials Research, Tohoku University, 2-1-1 Katahira, Sendai, 980-8577, Japan

[2] Materials Science and Engineering, Texas A&M University, College Station, TX 77843-3003, USA

[3] Department of Biomolecular Engineering, Graduate School of Engineering, Tohoku University, Sendai 980-8579, Japan

[4] Graduate School of Environmental Studies, Tohoku University, Sendai, 980-8579, Japan

[5] Department of Biomedical Engineering, Texas A&M University, College Station, TX 77843-3120, USA

[6] School of Computational Sciences, Korea Institute for Advanced Study, Seoul, 130-722, Korea

[7] Department of Physics and Astronomy, Texas A&M University, College Station, TX 77843-4242, USA

TEXT. Biological evolution has generated sophisticated and complex molecular motors, fitted for nanoscale transport.[1] Kinesin-1 (hereafter referred to as kinesin) carries out intracellular cargo transport along microtubule (MT) tracks. Kinesin is characterized by its processivity, i.e. its ability to detach of the track only after a relatively long traveled distance (typically several micrometers). In one second, kinesin is able to perform about 100 steps of 8 nm each, which gives an idealized speed of 800 nm/s.[2] The kinesin motor is fueled by one adenosin triphosphate (ATP) per step.[3] ATP is hydrolyzed into adenosin diphosphate and the resulting chemical energy is converted into mechanical work with a yield as high as 50%.[4,5] The force of translocation, ~6 pN,[6] is high enough to allow transportation of micrometer size cargo.[7] These specificities, unequaled by synthetic molecular motors,[8-10] make it a good candidate for nanotransport applications.[11-14] Kinesin can be used in two different ways for this purpose. The first way mimics the intracellular environment by using the MT as a track and kinesin as a cargo transporter.[15] The second way inverts the intracellular configuration by immobilizing kinesin proteins on the surface and making them propel the MTs, effectively acting like a conveyor belt.[2] Without constraints, the gliding direction of MTs is random. Hence, transport applications require guiding by patterns with selective absorption sites achieved physically,[16] chemically[17] or as a combination of both.[18] For example, MTs can be geometrically confined either in enclosed channels or thanks to an undercut, restraining any MT escape.[19,20] In each case, the width of the track is limited by lithographic fabrication techniques. Functionalized MTs can be used as a kinesin powered shuttle to carry cargos. Due to their length, MTs offer a longer cruising range and a much higher carrying capacity than kinesin motors. Several kinds of cargo have been studied, for example virus particles,[21] vesicles[22] or carbon nanotubes.[23] Nevertheless, to our knowledge, MT gliding behavior on non planar surfaces has barely been studied.[24] Previously, it has been shown that functionalized carbon nanotubes can be carried and displaced by microtubules.[23] In this report, we exploit the high aspect ratio and the nanoscale diameter of multi-walled carbon nanotubes (MWCNTs) and use them as tracks for kinesin powered MT shuttles. For this purpose, the MWCNTs have to be aligned and their attachment to the surface has to sustain the fluid exchange necessary for gliding assays. Besides, the MWCNTs have to be compatible with the kinesin motor protein, which is very sensitive to the surface chemistry.[25] For these reasons, we exploit the strong noncovalent binding between streptavidin and biotin.[26] MWCNTs functionalized with streptavidin are attached to a biotinylated surface which in turn allows for attachment of biotinylated kinesin.[27] MWCNT tracks present a large potential as they are almost chemically inert and possess excellent mechanical properties. As opposed to MTs, MWCNTs are stable and their observation is not limited by fluorescence degradation. Their lateral size, difficult to reach by lithography, allows the elaboration of very narrow tracks, comparable to the MT diameter. Moreover, the metallic character of the MWCNT is promising for applications as it facilitates their organization and may allow the electrical control of MTs or cargos, using MWCNTs as an electrode.

MWCNTs can be aligned using techniques such as molecular combing[28] or manipulation with an atomic force microscope (AFM).[29] Molecular combing involves drying and therefore prevents any protein manipulation. Alignment using an AFM tip is slow and cannot be performed on a large scale using a large amount of MWCNTs. We therefore utilize another technique, dielectrophoresis (DEP), which exploits the behavior of MWNCTs in an AC electric field that polarizes the MWCNTs and aligns them parallel to the field.[30] This technique is easy to apply, fast, reproducible and scalable. It has been shown that a single MWCNT can be aligned and attached to electrodes using this method,[31, 32] allowing downscaling. Moreover, the technique can be performed in water, making it compatible with protein functionalized MWCNTs.

Here, we present a MWCNT immobilization technique, the relevant streptavidin functionalization and the transport behavior of MTs propelled by kinesin tethered on aligned MWCNTs. The principle of the experiment is illustrated in Figure 1. First, aminosilane molecules are attached to a glass surface, to which polyethylene glycol (PEG) chains are covalently bound using the *N*-Hydroxysuccinimide (NHS) group at the end of the chain. The 4 nm PEG chain is nearly straight in an aqueous environment, allowing easy access to the biotin located at its extremity. After this step, streptavidin-MWCNT conjugates are immobilized on the surface.

MWCNTs (1 mg), already functionalized with carboxyl groups, were incubated with 1 mg of streptavidin (19 nmols) in 1 mL PBS buffer at 37°C for 1 hour. The amount of streptavidin absorbed on the surface of the MWCNTs was evaluated using a colorimetric experiment [33] (data not shown), analyzing the protein concentration in the supernatant before and after the conjugation. The reaction was efficient with 92% of the protein in solution absorbed, therefore 17.5 nmols of streptavidin was attached to the surface of the MWCNTs, which is consistent with previous works.[23] The available surface of 1 mg MWCNT is 0.3 m$^2$, covered by 17.5 nmols of streptavidin. This translates to about 28000 streptavidin molecules on a single MWCNT. Considering a square shaped streptavidin with a 5 nm edge and 0.79 µm$^2$ as an average surface of a single MWCNT, 88% of the surface was coated.[34] As a result, the streptavidin density on the MWCNT is assumed to be about 35000 molecules µm$^{-2}$. This value is approximately ten times higher than the mininum kinesin density required for gliding (about 4000 µm$^{-2}$).[2]

The MWCNTs alignment has been tested in different conditions: biotinylated surface, untreated surface and non functionalized MWCNTs. When there is no functionalization of either the glass surface or the MWCNT, upon turning off the electric field, the MWCNTs disorganize and spread in the buffer (movie S1 and S3) whereas they stay organized on the biotinylated surface (Figure 2, movie S2). This immobilization even resists drying and flow forces applied during fluid exchange. The flow cells have been dismantled and their surface dried in order to investigate the MWCNT organization by scanning electron microscopy (SEM). Two electrode configurations have been tested. In the first one, in order to connect the MWCNT tracks, the electrodes are shaped in triangles and crenellated (Figure 2, 3 a,b and S1). The second one consists of an interdigitated electrodes array (IDA) (Figure 3 c,d and S2). In both configurations, the images show aligned MWCNT bundles between electrodes. On average, the MWCNT are oriented orthogonally to the electrode surface (Figure S3), i.e. parallel to the electric field, as one would expect. The majority of segments are oriented within an angle of 30° from the electric field, although the alignment may have been better prior to the drying step necessary for the SEM characterization. Depending on the MWCNT concentration, the bridge between electrodes can be as thin as one MWCNT. The measured diameter is consistent with previous reported data showing a wide distribution (40-90 nm).[35]

The guiding capacity of the MWCNT tracks has been tested by performing gliding assays. First, biotinylated kinesin is inserted and bound to streptavidins coated MWCNTs. Rhodamine labeled MTs are then introduced and are observed by fluorescence microscopy. Figure 4 shows the gliding of MTs along a MWCNT bundle after casein passivation. In Figure 4 a), the MT is partially hidden as it glides on the side of the MWCNT (movie S4). In Figure 4 b), a MT moves on top of the bundle (movie S5). MTs can also glide in a different orientation from the bundles.

Several configurations can explain this. As the MT can be propelled by a single kinesin[2], a single attachment point on the bundle is sufficient to allow gliding, regardless of the MT orientation. It is also possible that the MT glides on an unaligned MWCNT or is attached on several aligned MWCNT.

The measured average gliding speed on the bundles is 149±53 nm/s (n=31). This velocity is comparable with values reported elsewhere[36, 37] but somewhat lower than the cargo speed motion previously measured using the same kinesin construct.[38] This can be explained by the different configuration of the assay. In the case of cargo transport by kinesin, a single motor protein is involved, ruling out the collective behavior issues occurring in MT gliding.[39-41] This collective behavior seems to depend on the kinesin's ability to twist in order to adapt to the MT direction.[42] This capacity is negatively affected by the truncation of the protein, decreasing its torsional flexibility and rigidifying the mechanical coupling between motors.[36] Moreover, our kinesin construct is truncated at residue 400, i.e. in the middle of the hinge located between residues 377 and 434, a part which has been shown to be important to achieve high gliding velocities.[43] Deletion, truncation or alteration has been reported to reduce the speed 4 to 5 fold. Considering a usual gliding velocity of 800 nm/s for a full-length kinesin, this is quantitavely consistent with our results.

Streptavidin functionalized MWCNT tracks have been elaborated using DEP and orthogonal chemistry. These MWCNT patterns show sufficient stability to sustain flow in the absence of electric field. This technique should be useful for future nanodevice applications in which high organization and stability are required. MT guiding and translocation along these tracks have been shown. The metallic character of these MWCNT tracks opens new possibilities, such as electrical control of MTs. Moreover, we can expect the emergence of single MWCNT tracks which will be useful for molecular transport devices.

**Materials and Methods.** Indium tin oxide (ITO) glass slides were provided by Sanyo Vacuum Industries Co., Ltd. (Tokyo, Japan). Hexamethyldisilazane was purchased from Tokyo Ohka Kogyo Co., Ltd. (Kanagawa, Japan). Positive g-line photoresist (S1818) and developer (MF CD-26) were obtained from Shipley Far East Ltd. (Tokyo, Japan). (3-Aminopropyl)triethoxysilane was obtained from Sigma-Aldrich Chemical Co. (St. Louis, MO, US).

MWCNT characterization and functionalization. Highly pure MWCNTs synthesized by a catalytic chemical vapor deposition process and graphitized/purified at approximately 3000°C were purchased from Hodogaya Chemical Co., Ltd. They were then characterized with a field-emission SEM (JSM-6500F, JEOL, Japan) and LaB6 gun-200KV TEM (JEM-2100, JEOL, Japan). The surfaces of the pristine hydrophobic MWCNTs were functionalized by a controlled acid treatment process as described elsewhere.[23] Briefly, MWCNTs were refluxed in 120 mL of 3:1 v/v ratio of 98% $H_2SO_4$ and 68% $HNO_3$ for 20 min at 110°C. The treated MWCNTs were then thoroughly washed with ultrapure water on a 1.2 μm membrane, sonicated, and dispersed in ultrapure water (~2 mg mL$^{-1}$) for several hours. The resulting MWCNTs were hydrophilic (zeta potential ~-40 mV in an aqueous solution with pH of ~4.1) with a high degree of crystallinity, as previously demonstrated by Raman spectroscopy.[35] The zeta potential of the acid-treated MWCNTs was measured by a Zetasizer Nano-Z (Malvern Instruments Ltd., UK) and calculated using the Smoluchowski equation. Raman spectra were recorded on a micro-Raman spectrometer (Horiba Jobin-Yvon T64000) using a 514.532 nm laser beam at room temperature.

MWCNTs were cleaned two times by centrifugation (15000 rpm, 3 min) in milli-Q water and diluted in phosphate buffered saline (PBS, pH 7.6) to a final concentration of 1 mg/mL.

MWCNTs were incubated for 1 hr at 37ºC with 1 mg mL$^{-1}$ of streptavidin in PBS buffer. After incubation, MWCNTs were washed again with PBS two times and the first supernatant was collected to determine the efficiency of the coupling. The efficiency of the coupling was evaluated by the Bradford test analyzing the protein concentration in the supernatant before and after the conjugation. The MWCNTs modified with streptavidin were stored at 4ºC in PBS buffer at 1 mg mL$^{-1}$ concentration.

Design and fabrication of the DEP devices. The ITO electrodes were fabricated on the glass slide by conventional photolithography and chemical etching using an etchant solution (HCl:H$_2$O:HNO$_3$/4:2:1 by volume) for 15 min under ultrasonication. The effective electrode dimensions were 1 × 1 mm$^2$ and 1 × 0.8 mm$^2$ for the crenellated and ITO-IDA devices, respectively (Figure S1 and S2). Hexamethyldisilazane and S1818 were poured onto the glass slide (thickness 1 mm; Matsunami Co., Japan) and it was baked at 90ºC for 10 min. It was then irradiated by UV light through the mask aligner (MA-20; Mikasa Co. Ltd., Tokyo, Japan), and developed with MF CD-26. After 15 min of sonication in the etching solution, the S1818 polymer was removed using acetone.

In the case of the ITO-IDA devices, an extra 20 µm thick insulating layer was formed on the electrode by SU-8 3020 polymer. Several 50 µm wide channels with 200 µm separation were formed on ITO by SU-8 3020.

Fabrication of aligned MWCNTs. The ITO electrodes were treated with plasma oxygen and silanized with (3-aminopropyl)triethoxysilane under vacuum for 1 hr. Further on, surfaces of the electrodes were treated with a 0.2 mg mL$^{-1}$ dilution of NHS-PEG$_{12}$-Biotin and NHS-PEG$_3$-Biotin in dimethyl sulfoxide solvent. After one hour of incubation at room temperature the electrodes were washed with ethanol and stored at 4ºC. A closed chamber was constructed to establish an AC electric field required for the DEP MWCNTs patterning. Polyethylene terephthalate (PET) film spacers (thickness, 35 µm) were used to define a chamber between the glass slide and the ITO electrodes. From the stock of functionalized MWCNTs different dilutions were prepared (0.1, 0.2, 0.5 mg mL$^{-1}$) in milli-Q water just prior to the experiment to avoid streptavidin damage. 20 µL of the MWCNTs were introduced in the microfluidic chamber.

The ITO electrodes were exposed to an AC voltage using a waveform generator (No. 7075, Hioki EE Co., Japan). An oscilloscope (Wave surfer 424; LeCroy Co., Japan) was used to confirm the generated electric current. The DEP-induced behavior of the MWCNTs was created and recorded by using an optical microscope (DMIRE2; Leica Co., Germany) equipped with a digital CCD camera (DFC350X; Leica Co., Germany). After 5 min the voltage was turned off and the fluid in the microfluidic chamber was replaced two times, first with milli-Q water and second with PEM buffer.

MT and kinesin. Rhodamine labeled MTs were polymerized from commercially available porcine tubulin (Cytoskeleton, USA) according to a protocol described elsewhere.[44] Truncated *Drosophila melanogaster* kinesin (400 residues, K400Bio), complemented with a Biotin Carboxyl Carrier Protein and a hexahistidine tag, was expressed and purified as described elsewhere.[45]

Motility assay. After MWCNT alignment, the chamber was cautiously rinsed with PEM buffer.[44] Then a casein solution (0.5 mg mL$^{-1}$ in PEM buffer) was introduced and incubated for 5 min. A kinesin solution (4 µM in PEM buffer) was then introduced and incubated for 10 min. Then a MT dilution (2.5 mg mL$^{-1}$) complemented with β-mercaptoethanol (0.5% v/v), antifade (20 µg mL$^{-1}$ Glucose Oxidase, 8 µg mL$^{-1}$ Catalase, 20 mM Glucose) and 2 mM ATP was added. Finally, the cell was sealed with VALAP (1:1:1 vaseline/lanolin/paraffin).

Fluorescence microscopy. The flow cell was mounted on an inverted microscope IX71 (Olympus, Japan) equipped with a digital CCD camera (Hanamatsu, ImageEM) and a rhodamine filter set (Omega Optical, Inc., XF204). Images were acquired and processed using Metamorph and ImageJ. MT tracking was performed using the MTrackJ plugin for ImageJ.

SEM. Samples were dried and coated with a thin layer of platinum using a JFC-1600 coater (JEOL, Japan). SEM images were acquired using a JSM-7800F Field Emission SEM (JEOL, Japan).

FIGURES.

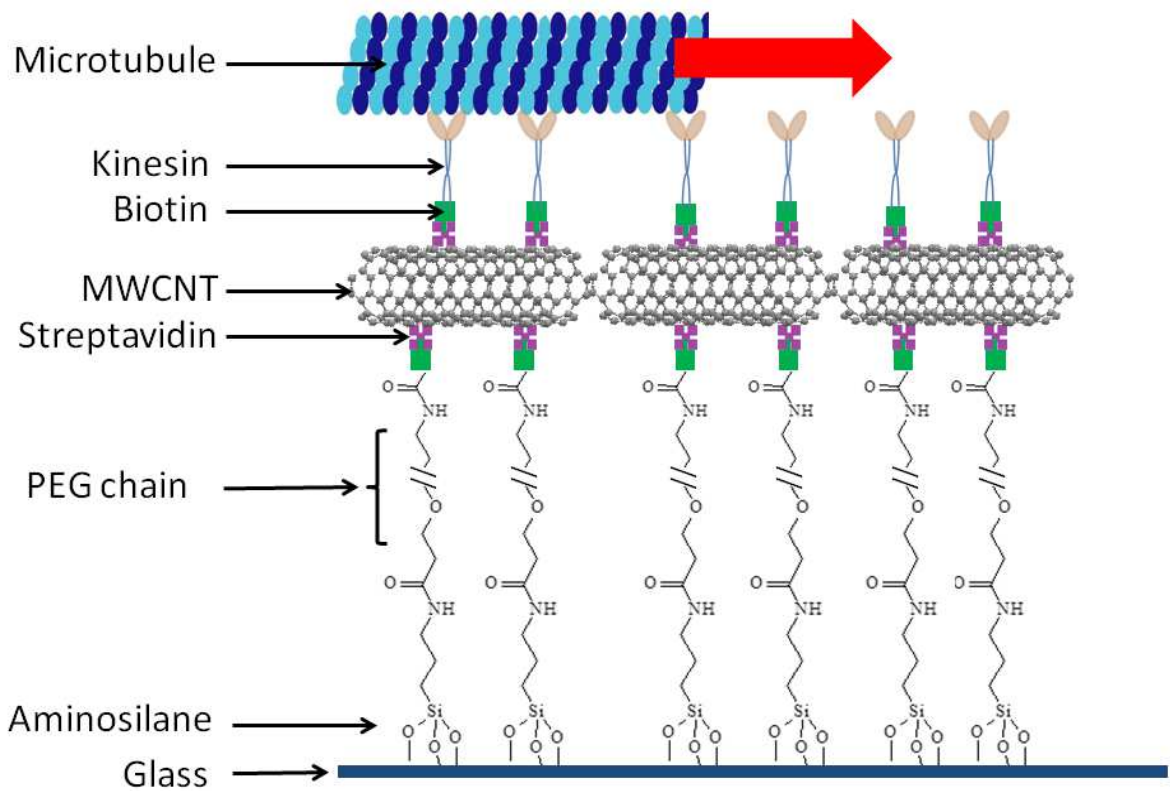

**Figure 1.** Schematic view of the experimental gliding system.

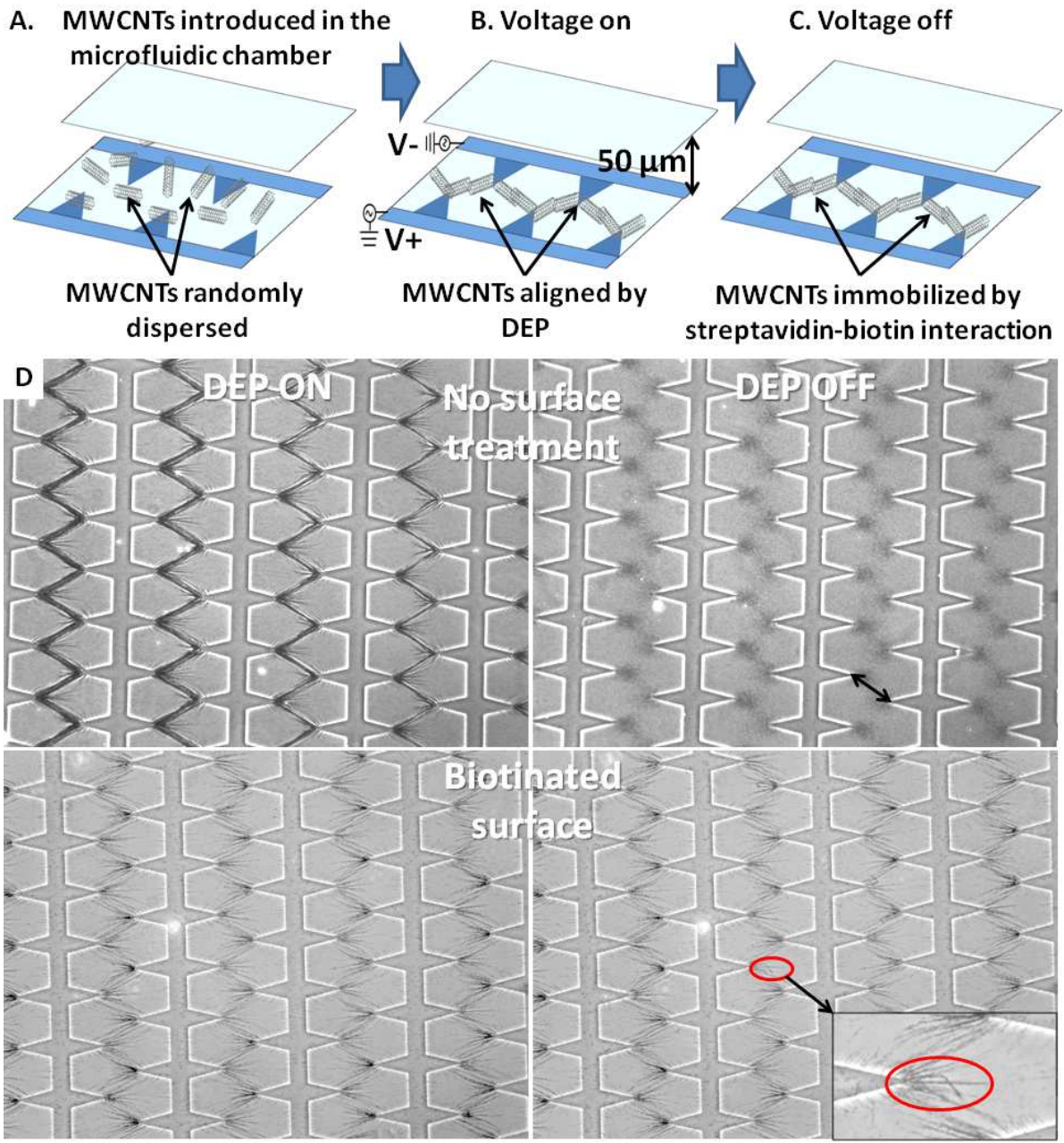

**Figure 2.** A,B,C) Schematic principle of the alignment by DEP. D) Bright field images of aligned MWCNT using DEP before and after surface treatment. On untreated surface, MWCNT disperse after turning off the voltage (movie S1). The space between two electrodes indicated by an arrow corresponds to 100 μm. The red area indicates the displacement of a bundle.

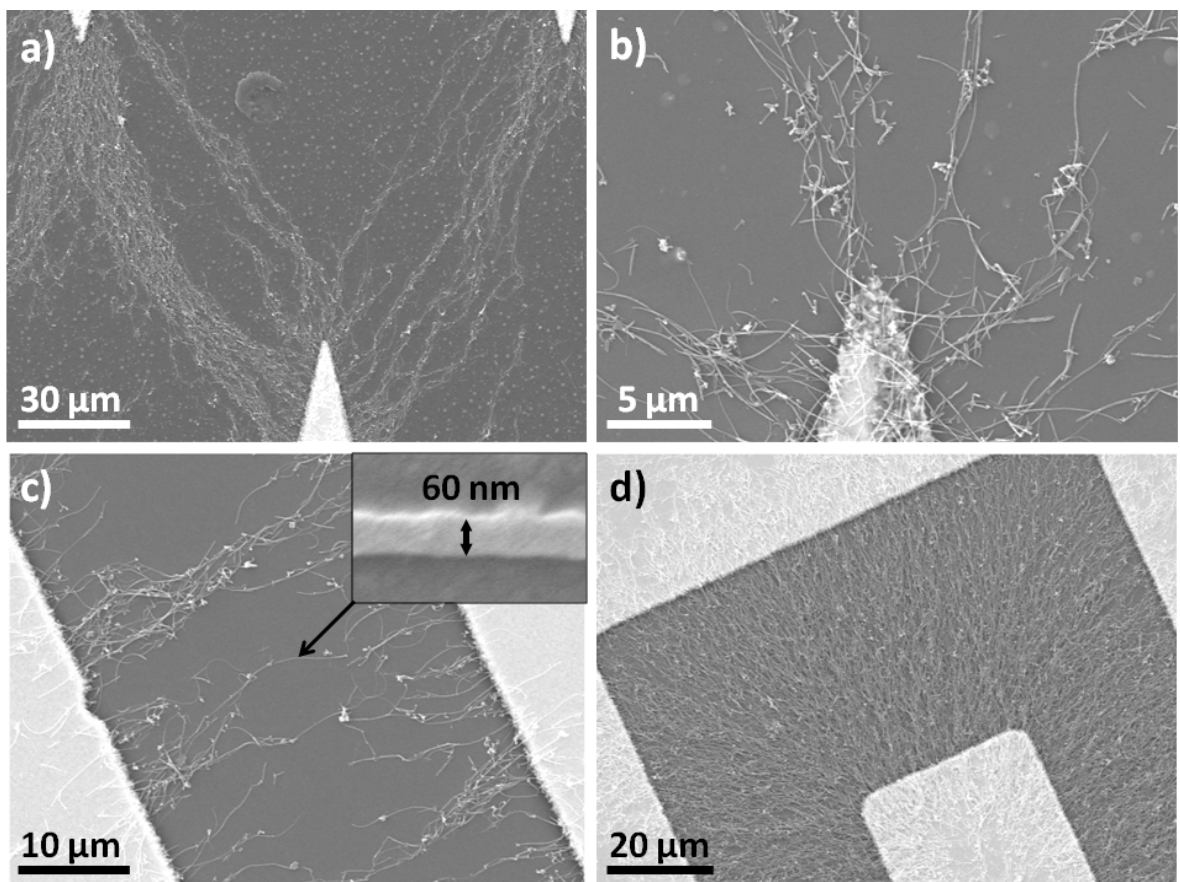

**Figure 3.** SEM images of aligned MWCNTs.

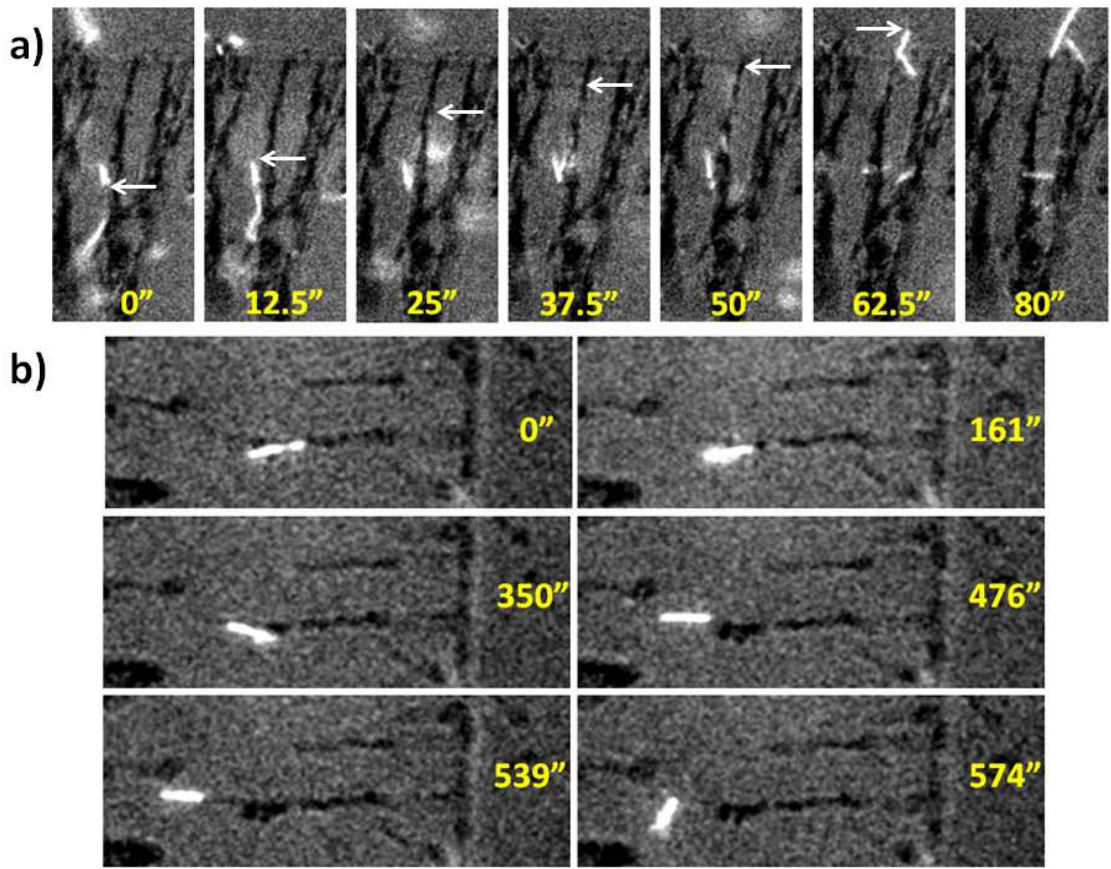

**Figure 4.** Bright field and fluorescence images of a MT gliding along a bundle of MWCNT. a) Gliding on the side of the bundle. The width of one image corresponds to 16.5 micrometers (movie S4). The white arrow indicates the top tip of the MT. b) Gliding on the top of the bundle. The width of one image corresponds to 29.4 micrometers (movie S5).

ASSOCIATED CONTENT

**Supporting Information**. Figures showing the ITO electrodes and resist patterns (Figure S1-S2). Histogram showing the angular repartition of the MWCNT segments (Figure S3). Kymographs extracted from the movies S4 and S5 (Figure S4). Figure illustrating the evolution of the MT trajectories and traveled distances (Figure S5). Real time movies showing the alignment of MWCNT by DEP in the case of non functionalized surface (Movie S1), biotinylated surface (Movie S2) and non treated MWCNT (Movie S3). Movies showing the motion of microtubules on MWCNT tracks (Movies S4-S5). This material is available free of charge via the Internet at http://pubs.acs.org.

AUTHOR INFORMATION

**Corresponding Author**

*E-mail: teizer@tamu.edu.

**Author Contributions**

A.S., J.R., M.U., H.S. and W.T. designed research. A.S., J.R., K.K., K.R. and H.N. performed research. A.S. analyzed data. A.S, J.R, W.H. and W.T. wrote the paper. I.K., T.A., T.M., W.H. and W.T. directed research. All authors have given approval to the final version of the manuscript.

**Notes**

The authors declare no competing financial interest.

ACKNOWLEDGMENT

We gratefully acknowledge support from the World Premier International Research Center Initiative (WPI), MEXT, Japan. We would like to thank Mr. Andrew L. Liao and Dr. Daniel Oliveira for their helpful contribution as well as Dr. Hideaki Sanada, in the Kumagai lab for generously providing the kinesin plasmid.